\documentclass[runningheads]{llncs}
\usepackage[T1]{fontenc}

\usepackage{graphicx}
\usepackage{svg}
\usepackage{multirow}
\usepackage{xcolor}
\usepackage{placeins}

\usepackage{enumitem}

\usepackage{siunitx}
\usepackage{stfloats} 
\usepackage{subcaption}
\usepackage{xspace}
\usepackage{circledsteps}
\usepackage{amsmath}
\usepackage{mathtools}
\usepackage[switch]{lineno}
\usepackage[hidelinks]{hyperref}

\usepackage{orcidlink}

\usepackage[ruled, linesnumbered]{algorithm2e}
\SetKwComment{Comment}{//}{}

\definecolor{lightblue}{HTML}{CCE5FF}
\usepackage{soul}
\sethlcolor{lightblue}
\soulregister\acrlong7
\soulregister\ac7
\soulregister\acf7
\soulregister\acs7
\soulregister\acp7
\soulregister\cite7
\soulregister\cref7
\soulregister\SI2
\soulregister\mathrm7
\soulregister\num7
\soulregister\volt7
\soulregister\$7
\soulregister\linelabel7

\usepackage{lineno}

\usepackage{tikz}
\usepackage[]{pgfplots}
\usepackage{pgfplotstable}
\pgfplotsset{compat=1.18}
\pgfplotsset{
    General/.style={
        font=\footnotesize,
        width=\linewidth,
        height=3.5cm,
        xtick pos=left,
        xtick align=outside,
        ytick pos=left,
        ytick align=outside,
        ymajorgrids=true,
        grid style=dashed,
        legend cell align={left},
        style=thin,
    },
    BarConfig/.style={
        General,
        ymin=0,
        enlarge x limits=0.25,
        xtick=data,
        xticklabel style={rotate=30, anchor=east},
        xlabel shift=-6pt,
        bar width=8pt,
    },
    ScatterConfig/.style={
      General,
      only marks,
      xtick=data,
      xticklabel style={rotate=90, anchor=east},
    }
}

\usepgfplotslibrary{statistics}
\makeatletter
\pgfplotsset{
    boxplot prepared from table/.code={
        \def\tikz@plot@handler{\pgfplotsplothandlerboxplotprepared}%
        \pgfplotsset{
            /pgfplots/boxplot prepared from table/.cd,
            #1,
        }
    },
    /pgfplots/boxplot prepared from table/.cd,
        table/.code={\pgfplotstablecopy{#1}\to\boxplot@datatable},
        row/.initial=0,
        make style readable from table/.style={
            #1/.code={
                \pgfplotstablegetelem{\pgfkeysvalueof{/pgfplots/boxplot prepared from table/row}}{##1}\of\boxplot@datatable
                \pgfplotsset{boxplot/#1/.expand once={\pgfplotsretval}}
            }
        },
        make style readable from table=lower whisker,
        make style readable from table=upper whisker,
        make style readable from table=lower quartile,
        make style readable from table=upper quartile,
        make style readable from table=median,
        make style readable from table=average,
        make style readable from table=lower notch,
        make style readable from table=upper notch,
        make style readable from table=draw position,
}
\makeatother

\usepackage[capitalise]{cleveref}
\usepackage[printonlyused,nolist]{acronym}
\def\BibTeX{{\rm B\kern-.05em{\sc i\kern-.025em b}\kern-.08em
    T\kern-.1667em\lower.7ex\hbox{E}\kern-.125emX}}

\usepackage{listings}
\usepackage{color} 

\usepackage{enumitem}

\definecolor{deepblue}{rgb}{0,0,0.5}
\definecolor{deepred}{rgb}{0.6,0,0}
\definecolor{deepgreen}{rgb}{0,0.5,0}
\definecolor{backcolour}{rgb}{0.95,0.95,0.92}

\crefname{lstlisting}{Listing}{Listings}
\Crefname{lstlisting}{Listing}{Listings}

\begin{acronym}
    \acro{rram}[RRAM]{Resistive Random Access Memory}
    \acro{tnn}[TNN]{Ternary Neural Network}
    \acro{cim}[CIM]{Computing-in-Memory}
    \acro{pim}[PIM]{Processing-in-Memory}
    \acro{mvm}[MVM]{Matrix-Vector Multiplication}
    \acro{mmm}[MMM]{Matrix-Matrix Multiplication}
    \acro{gemm}[GeMM]{General Matrix Multiply}
    \acro{mac}[MAC]{Multiply–Accumulate}
    \acro{nbb}[NBB]{NeuroBreakoutBoard}
    \acro{nvm}[NVM]{Non-Volatile Memory}
    \acro{tia}[TIA]{TransImpedance Amplifier}
    \acro{hrs}[HRS]{High Resistive State}
    \acro{lrs}[LRS]{Low Resistive State}
    \acro{wl}[WL]{Word Line}
    \acro{bl}[BL]{Bit Line}
    \acro{nn}[NN]{Neural Network}
    \acro{cnn}[CNN]{Convolutional Neural Network}
    \acro{sp}[SP]{Scalar Product}
    \acro{bnn}[BNN]{Binary Neural Network}
    \acro{qnn}[QNN]{Quantized Neural Network}
    \acro{adc}[ADC]{Analog-to-Digital Converter}
    \acro{dac}[DAC]{Digital-to-Analog Converter}
    \acro{api}[API]{Application Programming Interface}
    \acro{ispva}[ISPVA]{Incremental Step Pulse Program and Verify Algorithm}
    \acro{igvva}[IGVVA]{Incremental Gate Voltage and Verify Algorithm}
    \acro{igvp}[IGVP]{Incremental Gate Voltage Programming}
    \acro{dapv}[DAPV]{Dispersion-Aware Program-Verify}
    \acro{sp}[SP]{Single Pulse}
    \acro{nmos}[NMOS]{Negative channel Metal-Oxide Semiconductor}
    \acro{mos}[MOS]{Metal-Oxide Semiconductor}
    \acro{cmos}[CMOS]{Complementary Metal-Oxide Semiconductor}
    \acro{mim}[MIM]{Metal-Insulator-Metal}
    \acro{oxram}[OxRAM]{Oxide-based Random Access Memory}
    \acro{sl}[SL]{Source Line}
    \acro{c2c}[C2C]{Cycle-to-Cycle}
    \acro{d2d}[D2D]{Device-to-Device}
    \acro{cdf}[CDF]{Cumulative Distribution Function}
    \acro{cf}[CF]{Conductive Filament}
    \acro{wm}[WM]{Window Margin}
    \acro{if}[IF]{Incremental Form}
    \acro{ifv}[IFV]{Incremental Form and Verify}
    \acro{vcm}[VCM]{Valence Change Memory}
    \acro{uart}[UART]{Universal Asynchronous Receiver/Transmitter}
    \acro{protobuf}[Protobuf]{Protocol Buffer}
    \acro{trl}[TRL]{Technology Readiness Level}
    \acro{tvm}[TVM]{Tensor Virtual Machine}
    \acro{te}[TE]{Tensor Expression}
    \acro{ir}[IR]{Intermediate Representation}
    \acro{ifm}[IFM]{Input Feature Map}
    \acro{ofm}[OFM]{Output Feature Map}
    \acro{topi}[TOPI]{TVM Operator Inventory}
    \acro{1t1r}[1T1R]{One Transistor One Resistor}
    \acro{ste}[STE]{Straight Through Estimator}
    \acro{mlc}[MLC]{Multi-Level Cell}
    \acro{ibs}[IBS]{Input Bit Slicing}
    \acro{dse}[DSE]{Design Space Exploration}
    \acro{pcm}[PCM]{Phase Change Material}
    \acro{c2c}[C2C]{Cycle-to-Cycle}
    \acro{d2d}[D2D]{Device-to-Device}
\end{acronym}

\definecolor{c0}{HTML}{1b85b8}
\definecolor{c1}{HTML}{5a5255}
\definecolor{c2}{HTML}{559e83}
\definecolor{c3}{HTML}{ae5a41}
\definecolor{c4}{HTML}{c3cb71}
\definecolor{c5}{HTML}{bfb5b2}

\definecolor{cweighted}{HTML}{a72090}
\definecolor{cminerror}{HTML}{008080}

\usepackage[all]{background}
\usepackage{stackengine}
\setstackEOL{\\}
\setstackgap{L}{\normalbaselineskip}
\SetBgContents{\color{gray}{\tiny \Longstack{
    PREPRINT - Accepted at the 25th International Conference on Embedded Computer Systems: \\
    Architectures, Modeling and Simulation (SAMOS), June 28 - July 3, 2025}}}
\SetBgPosition{4.5cm,0.5cm}
\SetBgOpacity{1.0}
\SetBgAngle{0}
\SetBgScale{1.8}

\begin{document}

\renewcommand{\thelstlisting}{\arabic{lstlisting}}  
\setcounter{lstlisting}{0}

\title{
Optimizing Binary and Ternary Neural Network Inference on \acs{rram} Crossbars using CIM-Explorer
}
\titlerunning{CIM-Explorer}

\def\finalpaper{1}
\if\finalpaper1
\author{Rebecca Pelke\orcidlink{0000-0001-5156-7072} \and
José Cubero-Cascante\orcidlink{0000-0001-9575-0856} \and
Nils Bosbach\orcidlink{0000-0002-2284-949X} \and
Niklas Degener \and
Florian Idrizi \and
Lennart M. Reimann\orcidlink{0009-0003-5825-2665} \and
Jan Moritz Joseph\orcidlink{0000-0001-8669-1225} \and
Rainer Leupers}
\authorrunning{R. Pelke et al.}
\institute{{RWTH Aachen University, Germany}\\
\email{pelke@ice.rwth-aachen.de}}
\else
\author{Removed\\Removed}
\institute{{Removed}\\
\email{Removed}}
\fi

\maketitle
%
%
\begin{abstract}
Using \ac{rram} crossbars in \ac{cim} architectures offers a promising solution to overcome the von Neumann bottleneck.
Due to non-idealities like cell variability, \ac{rram} crossbars are often operated in binary mode, utilizing only two states: \ac{lrs} and \ac{hrs}.
\acp{bnn} and \acp{tnn} are well-suited for this hardware due to their efficient mapping.

Existing software projects for RRAM-based CIM typically focus on only one aspect: compilation, simulation, or \ac{dse}.
Moreover, they often rely on classical \SI{8}{\bit} quantization.

To address these limitations, we introduce \textit{CIM-Explorer}, a modular toolkit for optimizing \ac{bnn} and \ac{tnn} inference on \ac{rram} crossbars.
CIM-Explorer includes an end-to-end compiler stack, multiple mapping options, and simulators, enabling a \ac{dse} flow for accuracy estimation across different crossbar parameters and mappings.
CIM-Explorer can accompany the entire design process, from early accuracy estimation for specific crossbar parameters,
to selecting an appropriate mapping, and compiling \acp{bnn} and \acp{tnn} for a finalized crossbar chip.
In \ac{dse} case studies, we demonstrate the expected accuracy for various mappings and crossbar parameters.

\textit{CIM-Explorer can be found on GitHub\footnote{
    CIM-Explorer: \url{https://github.com/rpelke/CIM-E} \\
    Crossbar simulator: \url{https://github.com/rpelke/analog-cim-sim}
    }.
}
\end{abstract}


\acresetall 

\vspace{-0.2cm}
\section{Introduction}

\ac{cim} addresses the von Neumann bottleneck by fusing computation and storage.
\ac{rram} is a promising technology for \ac{cim} due to its energy efficiency, high device density, and \acs{cmos} compatibility~\cite{wu2022a,zahoor2020resistive}.
Using \ac{rram} for analog \ac{cim} introduces challenges such as \ac{c2c} and \ac{d2d} variability, thermal instability, limited endurance, and read disturb effects~\cite{pelke2024fully,pelke2025show,shim2020impact,swaidan2019rram}. 
Analog operations further suffer from input/output noise, wire resistance, and nonlinear device I-V characteristics, which are particularly problematic for \ac{mlc} \ac{rram}~\cite{ding2023bnn}.
In contrast, binary \ac{rram} is a simpler, more robust option.
Binary crossbars use two states per cell, called \ac{hrs} and \ac{lrs}.
Therefore, \acp{bnn} are well-suited for efficient mapping to binary crossbars~\cite{ding2023bnn,kim2018neural,sun2018xnor}.
The same applies to \acp{tnn}~\cite{morell2022ternary}, although \acp{tnn} are less common in previous works related to \ac{rram} crossbars.

To evaluate the potential of \acp{bnn} and \acp{tnn} for \ac{rram} crossbars at an early design stage, a variety of tools are required, including compilers, simulators, and \ac{dse} methods.
Many individual tools already exist in the \ac{cim} research field~\cite{chen2023autodcim,drebes2020tc,he2019noise,khan2022cinm,neurosim,rasch2021flexible,siemieniuk2021occ,CrossSim,vadivel2020tdo,zhu2023mnsim}.
However, these tools mainly focus on \SI{8}{\bit} or \SI{16}{\bit} workloads and only cover individual parts such as compilation or \ac{dse}.
There is no one-fits-all solution for \acp{bnn} and \acp{tnn}.

We close this gap by introducing CIM-Explorer, a comprehensive toolkit for exploring \ac{bnn} and \ac{tnn} inference on \ac{rram} crossbars.
CIM-Explorer is designed to support the entire design workflow, ranging from early-stage accuracy estimations to code generation for fabricated crossbar chips.
\Cref{fig:overviewshort} illustrates the toolkit's individual components.
We highlight the following contributions:

\begin{figure}[!tbp]
    \centering
    \includegraphics[width=\linewidth, keepaspectratio]{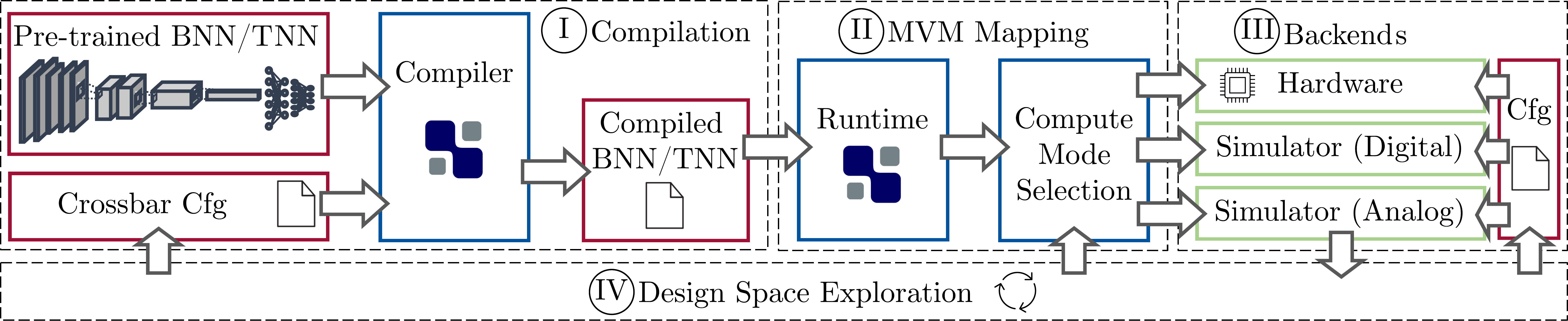}
    \caption{Overview of the individual modules of CIM-Explorer.
    At compile time, the \ac{bnn} or \ac{tnn} is optimized for execution on a crossbar \Circled{I}.
    During runtime, the weights are prepared according to the compute mode \Circled{II}.
    Several backends can be used for execution, e.g. different simulators \Circled{III}.
    A \ac{dse} tool automates finding optimal crossbar parameters and mappings \Circled{IV}.
    }
    \label{fig:overviewshort}
    \vspace{-0.4cm}
\end{figure}

\begin{itemize}
    \item[\Circled{I}] A \textbf{\ac{tvm}-based compiler} including a new Larq~\cite{geiger2020larq} frontend, multi-batch support, and crossbar-specific optimizations, e.g., maximizing weight reuse.
    It supports arbitrary crossbar sizes.
    Larq is an open-source training framework for \acp{bnn} and \acp{tnn}.
    \item[\Circled{II}] The implementation of different mapping techniques, also called \textbf{compute modes} in the following. The compute modes differ in, e.g., the handling and interpretation of negative inputs and weights.
    \item[\Circled{III}] Well-defined interfaces so that different types of \textbf{simulators} or even real hardware (if available) can be used as a target for execution.
    \item[\Circled{IV}] A \textbf{\ac{dse} flow} that uses the components \Circled{I-III} to analyze the impact of crossbar parameters, \ac{adc} parameters, and compute modes on the inference accuracy.
\end{itemize}

\Cref{sec:relatedwork} provides a comparison to existing \ac{cim} compilers and \ac{dse} frameworks.
\Cref{sec:background} presents all relevant background information regarding \acp{bnn}, \acp{tnn}, and \ac{rram}.
\Cref{sec:implementation} focuses on the implementation, followed by a \ac{dse} in \Cref{sec:results}, and a conclusion in \Cref{sec:conclusion}.
While this work focuses on accuracy as a metric, there is an extension that analyzes energy efficiency~\cite{cubero2025evaluating}.

\vspace{-0.3cm}
\section{Related Work}
\label{sec:relatedwork}
\vspace{-0.2cm}
We categorize this section into \ac{dse} approaches and compilers for \ac{cim} targets.
These areas have often been treated separately in previous research.
Our work combines these topics because the execution order determined during compilation can affect accuracy on non-ideal hardware.
Using separate tools for DSE and compilation can lead to discrepancies between simulated and real results due to differing execution orders.
Our integrated approach eliminates this risk by ensuring that the \ac{dse} process utilizes the same compiler, maintaining consistency between \ac{dse} and code generation for real hardware.

\vspace{-0.2cm}
\subsection{DSE Tools}
\label{sec:dse_tools_relwork}
Existing \ac{dse} frameworks differ in abstraction level (architecture or crossbar level) and focus (training or inference).
Established open-source frameworks include NeuroSim~\cite{neurosim},
MNSIM~\cite{zhu2023mnsim}, Aihwkit~\cite{rasch2021flexible}, PytorX~\cite{he2019noise}, and CrossSim~\cite{CrossSim},
which will be introduced in the following.

\textbf{NeuroSim} is an end-to-end benchmarking framework for \ac{cim} accelerators, including device-to-algorithm-level design options.
Besides evaluating the inference accuracy for various \ac{cim} technologies, NeuroSim also assesses the entire chip-level architecture.
However, it only focuses on \SI{8}{\bit} inputs and weights.

\textbf{MNSIM 2.0} is a behavior-level simulator for \ac{pim} architectures.
It provides a hierarchical modeling structure for both digital and analog \ac{pim}, supporting \ac{nn} accuracy estimation and a \ac{pim}-oriented \ac{nn} model training and quantization flow.
It focuses on the architecture level rather than the crossbar level.
Similar to NeuroSim, it uses custom layer descriptions for quantization, which makes adding new applications cumbersome.

\textbf{Aihwkit} is an open-source, PyTorch-based toolkit for simulation, training, and inference on analog crossbar arrays.
It focuses on the concept of an \textit{analog tile} for building \acp{nn} with analog components, allowing emulation of various hardware characteristics and non-idealities.
It focuses on analog hardware-aware training of floating-point models.
Using pre-trained models is not possible.

\textbf{PytorX} is a PyTorch-based toolkit for fault-aware training and inference on \ac{rram} crossbar arrays.
This framework profiles the behavior of the crossbar and injects noise into the training process to improve classification accuracy.
However, their workloads are limited to \SI{8}{\bit} \acp{nn}.

\textbf{CrossSim} is a GPU-accelerated framework developed for simulating \ac{nn} inference on analog \ac{cim} accelerators.
Similar to our \ac{dse}, it focuses on the impacts of hardware non-idealities on accuracy.
CrossSim allows for detailed configuration of \ac{nn} models, quantization parameters, and hardware characteristics, including device variability and non-idealities.
It is optimized for performance but cannot take resource constraints like a limited number of crossbars into account.
In addition, the focus is on \SI{8}{\bit} workloads.

\subsection{CIM Compilers}
\label{sec:relworkcompiler}
\begin{figure}[!tp]
    \centering
    \includegraphics[width=.6\linewidth, keepaspectratio]{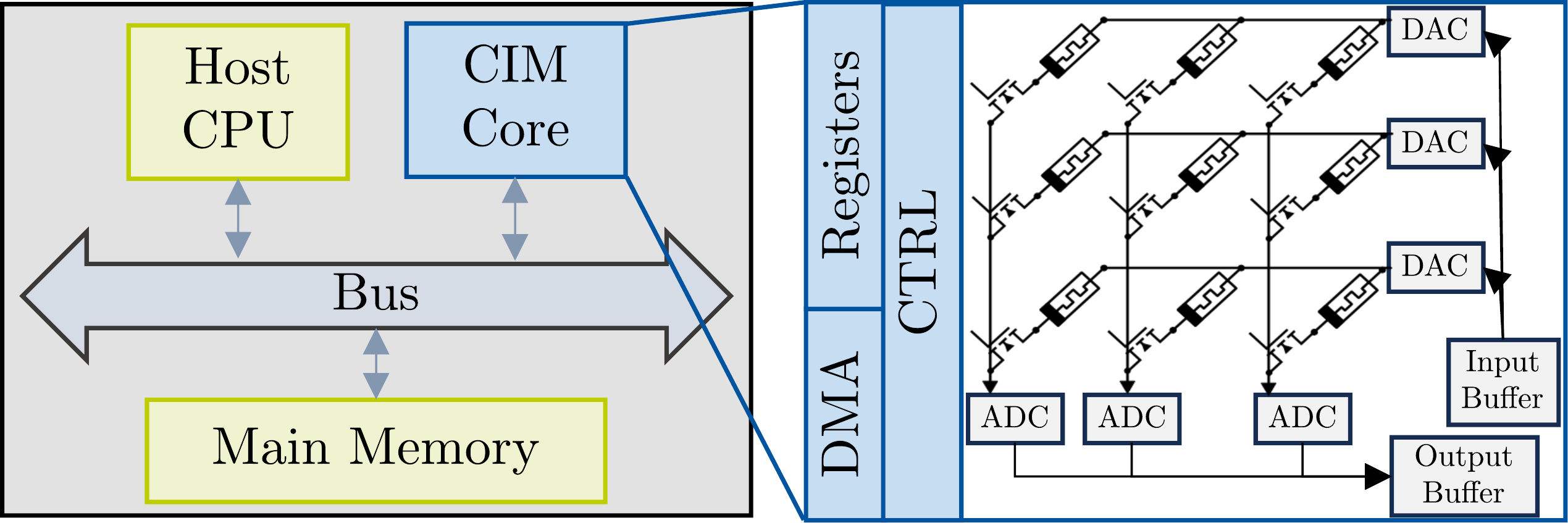}
    \caption{The CIM architecture components considered in this work.}
    \label{fig:architecture}
    \vspace{-0.4cm}
\end{figure}

Many compilers for \ac{cim} targets are available~\cite{chen2023autodcim,khan2022cinm}.
They differ in the target architecture, implemented optimizations, and offloaded patterns (e.g., \ac{mvm}, \ac{gemm}).
In the following, we focus on TC-CIM~\cite{drebes2020tc}, TDO-CIM~\cite{vadivel2020tdo}, and OCC~\cite{siemieniuk2021occ} since their view of the target architecture is similar to ours.
\Cref{fig:architecture} illustrates our view of the system architecture containing the \ac{cim} accelerator with its memory-mapped interface.
The \ac{nn} inference starts on the CPU.
The selected \ac{cim} patterns, in our case \acp{mvm}, are offloaded to the accelerator.

\textbf{TC-CIM}~\cite{drebes2020tc} uses Tensor Comprehensions~\cite{vasilache2019next} and Loop Tactics~\cite{chelini2019declarative} to detect and offload suitable tensor operations to a \ac{cim} accelerator.
After polyhedral optimizations with Tensor Comprehensions, Loop Tactics detects patterns like \ac{mvm}, \ac{gemm}, or batched \ac{gemm}.
The compiler is validated using a Gem5 simulator, including a \SI{4}{\bit} $256{\times}256$ \ac{pcm} crossbar.

The compilation approach of \textbf{TDO-CIM}~\cite{vadivel2020tdo} is similar to the one used in TC-CIM.
However, in TDO-CIM, the pattern recognition with Loop Tactics is done on LLVM-IR. The input of TDO-CIM is C/C++ code.
As in TC-CIM, only individual layers are simulated, and not entire \acp{nn}.

\textbf{OCC}~\cite{siemieniuk2021occ} uses MLIR to offload \ac{gemm} operations to a \ac{cim} accelerator.
It transitions from the Linalg dialect to a \ac{cim}-specific dialect.
It includes hardware optimizations to fit computations within constrained crossbar sizes and to minimize the number of write operations.
The \ac{gemm} computations are replaced by function calls to the accelerator.
OCC addresses the limited endurance of the \ac{pcm} cells by minimizing the number of write operations to the crossbar.
This strategy enhances weight reuse and significantly increases the system's lifetime.

Unlike OCC, our compiler reduces the number of write operations not only across individual layers but also across multiple input batches.
Moreover, the previous compilers only support \SI{8}{\bit} or \SI{16}{\bit} workloads, and \ac{nn} accuracy is not evaluated.
With our compiler, entire \acp{nn} can be executed, allowing for the evaluation of classification accuracy in the context of crossbar inaccuracies.

\vspace{-0.3cm}
\section{Background}
\label{sec:background}
\vspace{-0.3cm}
This section presents the background related to \acp{bnn} and \acp{tnn}, and explains the basic concepts and notations regarding \ac{rram} crossbars.

\subsection{Binary and Ternary Neural Networks}
\label{sec:bnns}

In \acp{bnn} and \acp{tnn}, weights and activations are represented using only two or three states, respectively~\cite{li2016ternary,samiee2019low}.
They still maintain reasonable accuracy on standard datasets~\cite{courbariaux2016binarized}.
The \textit{sign} function is widely used for \ac{bnn} quantization~\cite{qin2020binary}, while the \textit{ternary} function is used for \ac{tnn} quantization ~\cite{li2016ternary}:
\vspace{-0.5cm}

\begin{equation}
sign(x) =
    \begin{cases}
        +1, &if \hspace{0.2cm} x \geq 0 \\
        -1, &otherwise
    \end{cases}
\\ \quad
ternary(x) =
    \begin{cases}
        +1, &if \hspace{0.2cm} x > \Delta \\
         0, &if \hspace{0.1cm} \left|x\right| \leq \Delta \\
        -1, &if \hspace{0.2cm} x < - \Delta
    \end{cases}
\end{equation}

The threshold $\Delta$ depends on the weights~\cite{li2016ternary}.
To train \acp{bnn} and \acp{tnn}, Larq can be used.
Larq is an open-source Python library that is built on top of TensorFlow's Keras.
It provides specialized optimizers, training metrics, and a model zoo with pre-trained models.~\cite{geiger2020larq}

\vspace{-0.2cm}
\subsection{Memristive Devices and \ac{rram} Crossbars}
\label{sec:rrambackground}
A memristive device is composed of a transition-metal-oxide layer between two conducting electrodes~\cite{hazra2021optimization}.
Before usage, each device must be \textit{formed}.
This process enables the resistance-switching behavior~\cite{poehls2021review}.
It impacts the lifespan, forming yield, and \ac{c2c} variability of the cell.
After forming, \textit{set} and \textit{reset} operations can be applied, which bring the device to the \ac{lrs} and \ac{hrs}~\cite{puglisi2014study}.
We will focus on \ac{1t1r} cells.
They consist of one transistor and one memristive device.
The transistor is used to disconnect the memristor from the crossbar, which reduces sneak-path currents~\cite{mao2016optimizing}.

\ac{rram} devices, a specific type of memristors, can be arranged in crossbar structures to facilitate in-memory computing,
i.e., executing \acp{mvm} in the analog domain~\cite{cao2021neural,pelke2023mapping}.
When conducting an \ac{mvm} operation, the conductance values of the crossbar cells represent the matrix.
The read voltages represent the input (vector), and the output currents correspond to the result of the \ac{mvm}.
An \ac{adc} converts the output current back into a digital value.

\vspace{-0.2cm}
\section{CIM-Explorer Implementation}
\label{sec:implementation}
\vspace{-0.2cm}

\begin{figure}[!tbp]
    \centering
    \includegraphics[width=\linewidth, keepaspectratio]{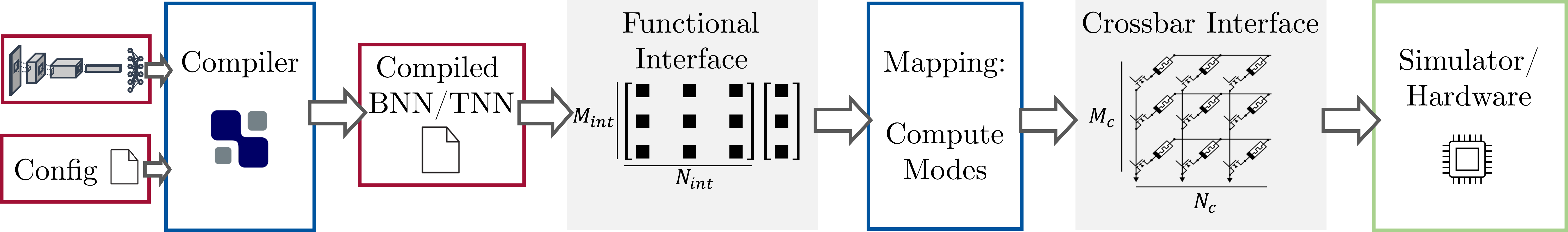}
    \caption{The interfaces of the toolkit.
    The functional interface separates compilation and mapping.
    The crossbar interface separates mapping and simulation.
    }
    \label{fig:overview_sw}
    \vspace{-0.4cm}
\end{figure}

This section provides an overview of the individual components.
The design goal is high modularity, allowing for the easy replacement of the \ac{nn}, mapping technique, and (simulator) backend.
To achieve this, we have defined interfaces between the individual components.
\Cref{fig:overview_sw} shows an overview of the interfaces.

The \textbf{functional interface} abstracts hardware-specific details from the compiler.
Typical values are, e.g., $M_{\text{c}}=N_{\text{c}}\in\{64, 128, 256, 512\}$~\cite{cubero2025evaluating}.
The compiler transforms the \ac{nn} to contain \acp{mvm} with a matrix dimension of $M_{\text{int}}{\times}N_{\text{int}}$,
which does not necessarily correspond to the size of the physical crossbar $M_{\text{c}}{\times}N_{\text{c}}$.
In many mappings, multiple \ac{rram} cells are used per weight.  
The compiler replaces the transformed \acp{mvm} with function calls.  
These calls remain unresolved during the compilation phase and are later resolved by the dynamic linker during inference.  
For inference, the functional interface must be implemented by a shared library that is loaded at runtime.

\begin{lstlisting}[language=C, 
    caption=Extract of the functional interface functions implemented in C.,
    label=lst:CfunctionCalls,
    floatplacement=H,
    numbers=left,
    aboveskip=0.5cm,   % Optional: Mehr Abstand oben
]
// Copy m_int x n_int matrix m to crossbar
// Layout of m: n-dimension first
int write_matrix(int *m, int m_int, int n_int);

// Execute MVM operation: r = m * v, r=result, v=vector
// Mapper knows matrix m from the previous write_matrix
int mvm(int *r, int *v, int m_int, int n_int);
\end{lstlisting}

\Cref{lst:CfunctionCalls} shows the function calls that must be implemented by the shared library.  
The functions pass pointers to vectors or matrices along with their dimensions.  
A row-major matrix layout is used.  
Separating write operations (\texttt{write\_matrix}) from the compute operation (\texttt{mvm}) allows the reuse of one matrix across multiple \acp{mvm}, thereby extending the lifespan of the cells.  
The toolkit includes two different libraries that implement the functional interface.  
One is written in C/C++ aims at fast simulation.  
The other library generates Python callbacks to simplify the initial prototyping of crossbar-specific features.  

The \textbf{crossbar interface} defines the interaction between the mapper and the hardware or simulator.  
It contains functions similar to those in \Cref{lst:CfunctionCalls} but with slightly different parameters.  
The mapper translates integer representations into the analog domain, while the simulator operates strictly on these analog values.  
As previously explained, both interfaces handle different matrix dimensions.  
The mapper resolves the relationship between $M_{\text{int}}{\times}N_{\text{int}}$ and $M_{\text{c}}{\times}N_{\text{c}}$ and invokes the crossbar interface to perform the actual computation of the \ac{mvm}.
Further details regarding the mapping are provided in \Cref{sec:intvscrossbararith}.  

\vspace{-0.3cm}
\subsection{TVM Compiler}
\label{sec:compiler}
\vspace{-0.2cm}

\ac{tvm} is a compiler framework that deploys deep learning models on a variety of hardware backends.
It is designed for CPUs, GPUs, and specialized accelerators~\cite{chen2018tvm}.
\ac{tvm} adopts the idea from Halide~\cite{ragan2017halide} of decoupling \textit{compute} and \textit{schedule},
meaning that for each compute definition, different schedules (implementations) can be selected.
\ac{tvm}'s \ac{te} concept allows to define those compute definitions.
To optimize loops, a schedule is built by progressively applying transformations,
known as \textit{schedule primitives},
which maintain the program's logical equivalence.
\ac{tvm} automatically generates its low-level representation, called TensorIR, from the schedule by applying four standard lowering phases.
Developers can insert custom passes after each phase.

\begin{figure*}[t!]
\centering
\includegraphics[width=\linewidth, keepaspectratio]{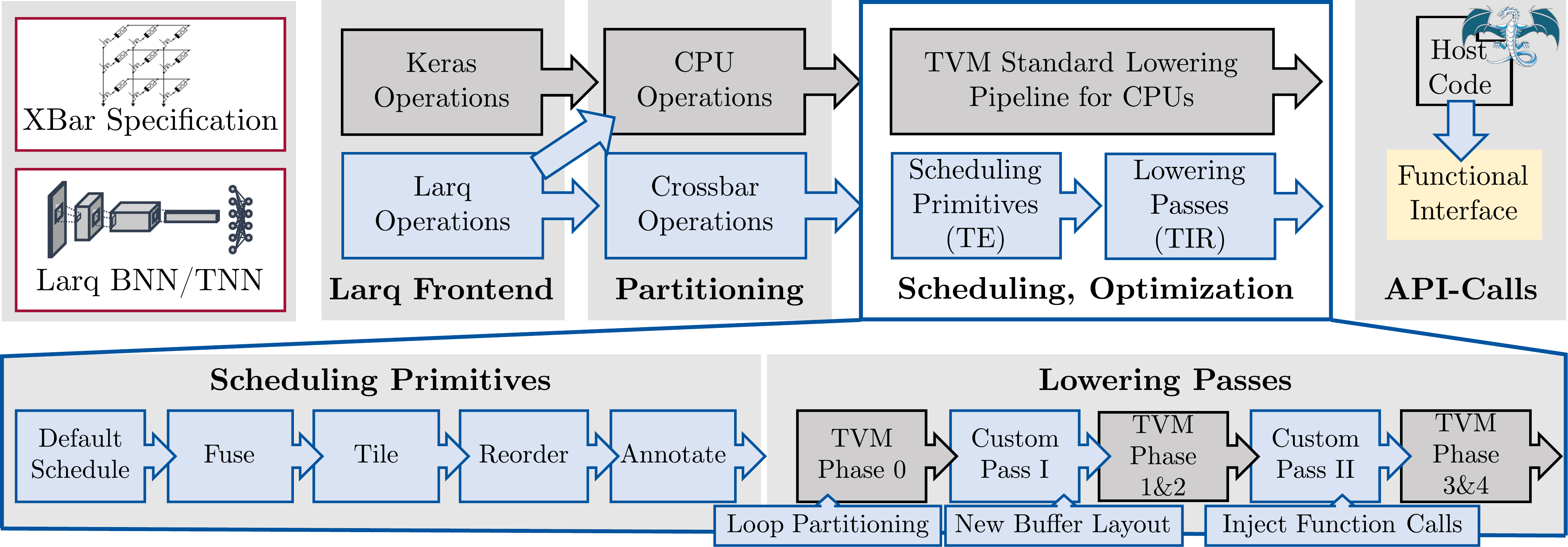}
\vspace{-0.6cm}
\caption{The compiler pipeline including pre-trained inputs, a new frontend, partitioning, scheduling primitives and lowering passes, and code generation.}
\label{fig:overview}
\vspace{-0.6cm}
\end{figure*}

\Cref{fig:overview} illustrates the developed compiler pipeline, which is used to convert the layers into \acp{mvm} of the required dimensions and offload the \acp{mvm} to the functional interface.  
Custom steps that differ from the standard \ac{tvm} pipeline are highlighted in blue.  
After defining hardware-specific properties, such as the crossbar dimensions $M_{\text{int}}{\times}N_{\text{int}}$,
a pre-trained Larq \ac{nn} is translated into the \ac{tvm}-specific high-level graph description called \textit{Relay}.  
Since Larq inputs are not supported in mainline \ac{tvm}, we developed the Larq frontend from scratch.  
This is done by expressing Larq-specific layers, such as \texttt{QuantDense} and \texttt{QuantConv2D}, through existing Relay operations.  
For all other layers, the standard \ac{tvm} pipeline for Keras or TensorFlow can be used.  
After building a Relay graph, the \ac{nn} operations are partitioned into CPU and crossbar operations using so-called \textit{Strategies}.  
A Strategy is a mechanism that allows developers to select different compute operations and schedules for the same operation depending on the target architecture.  
For the Conv2D operation, for example, we selected the standard \texttt{topi.nn.conv2d\_nhwc} operation as the compute operation and wrote a custom schedule for it, enabling the integration of function calls to the functional interface at a later stage.  
This custom schedule can be generated using \textit{scheduling primitives}.  
Finally, we implemented custom \textit{lowering passes} to inject \ac{api} calls.

\vspace{-0.5cm}
\subsubsection{Scheduling:}
\begin{figure*}[t!]
\centering
\includegraphics[width=\linewidth, keepaspectratio]{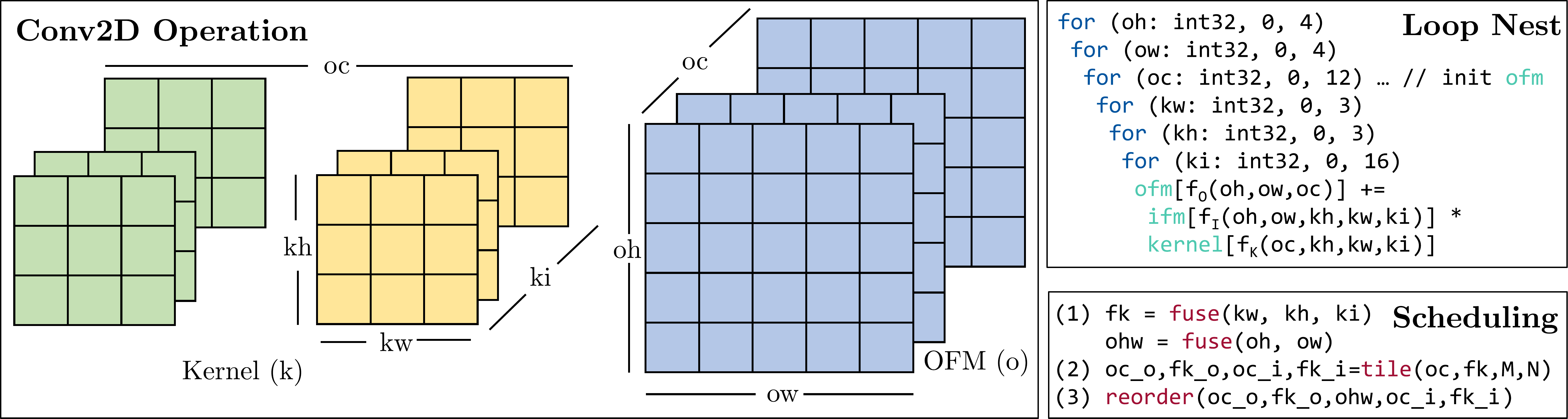}
\vspace{-0.6cm}
\caption{Scheduling primitives are applied to the initial loop nest of Conv2D.}
\label{fig:conv2d}
\vspace{-0.6cm}
\end{figure*}

\Cref{fig:conv2d} presents a simplified example for the loop transformations applied to a single-batch Conv2D operation.
The variables $kh, kw, ki$ and $oh, ow, oc$ refer to the loop axes, with $kx \in [0, K_X]$ and $ox \in [0, O_X]$.
The resulting loop nest contains six \texttt{for} loops.
The multidimensional tensor indices of the \ac{ifm}, kernel, and \ac{ofm} are simplified
as $f_\text{O}$, $f_\text{I}$, and $f_\text{K}$, respectively.
Scheduling primitives are applied to the loop nest to isolate the \ac{mvm} operation into the innermost loops and replace them with function calls to the functional interface (see \Cref{fig:overview_sw}).
This is achieved through the \texttt{reorder} primitive that reorders the axes after \texttt{tiling}. 
In practical terms, this process involves unrolling the kernels and grouping them into a matrix, as in im2col~\cite{yanai2016efficient}.
From this matrix, submatrices of size $M_{\text{int}}{\times}N_{\text{int}}$ are extracted.
Each submatrix is programmed into the crossbar (once) and used with various input vectors.
Finally, the CPU accumulates the partial results.

\vspace{-0.5cm}
\subsubsection{Lowering Passes:}
Replacing computations by function calls at the \ac{te} level is usually facilitated by \ac{tvm}'s concept called \texttt{tensorize}.
This concept requires the following conditions to be met: $K_\text{O}{\bmod}M~=~0$ and $\left(K_\text{H} K_\text{W} K_\text{I} \right){\bmod}~N~=~0$.
Because these conditions are not always met, \textit{if}-statements occur in the loop body to handle edge cases.  
These \textit{if}-statements cannot be handled by \texttt{tensorize}.  
To address this limitation, we perform loop partitioning after lowering phase~$0$.  
This replaces \textit{if}-statements by generating multiple loops at the same depth with different ranges.  
Additionally, we incorporate two custom lowering passes (see \Cref{fig:overview}):  
We add buffers of size $M_{\text{int}}{\times}N_{\text{int}}$, $1{\times}N_{\text{int}}$, and $1{\times}M_{\text{int}}$ to copy the values of the kernel, \ac{ifm}, and \ac{ofm}, respectively.  
This aligns the memory layout with the required format for calls to the functional interface.  
Then, we inject function calls that replace the isolated computation in the inner loop nest.  
Pointers to the buffers are passed as arguments to the function calls.

\vspace{-0.4cm}
\subsection{Integer-Crossbar Mapping}
\label{sec:intvscrossbararith}
\vspace{-0.2cm}
So far, we explained how to transform a pre-trained \ac{nn} and insert function calls of the functional interface.
The \ac{mvm} arithmetic of the functional interface is referred to as \textit{integer} arithmetic.
An \ac{mvm} in integer arithmetic cannot be executed directly on the crossbar due to the following reasons and assumptions:
\begin{itemize}[topsep=1pt]
\item Negative weights cannot be represented as negative conductance.
\item A conductance of \SI{0}{\siemens} (infinitely high resistance) cannot be achieved.
\item Read voltages are binary and only have one polarity, e.g., $V_\text{r}{\in}\{\SI{0}{\volt}, \SI{0.2}{\volt}\}$.
\end{itemize}

To handle negative weights, two approaches exist: \textit{linear-scaling mode} and \textit{differential mode}.
Linear scaling modifies the weights by scaling and adding an offset to achieve positivity~\cite{shafiee2016isaac}.
Differential mode employs two \ac{rram} cells for each integer value, with one representing the positive and one the negative part~\cite{fouda2019mask}.
Differential mappings reduce sensitivity to various categories of analog errors, including state-independent errors, state-proportional errors, and quantization errors~\cite{xiao2021accuracy}.
In contrast, linear-scaling mappings reduce the number of needed cells per weight, e.g., only one cell per weight for \acp{bnn}.

In the following sections, we will explain the different mapping options for \acp{bnn} and \acp{tnn} in more detail.
Each mapping is either based on the differential or linear-scaling idea.
First, we convert \acp{mvm}, so they only contain zeros and ones.
We call this \textit{digital} crossbar arithmetic (subscript \textit{D}).
This resolves all negative weights and inputs.  
In the final step, the digital \ac{mvm} is converted to voltages, conductances, and currents.
We call this \textit{analog} crossbar arithmetic (subscript \textit{A}).
The analog-crossbar-based \ac{mvm} is then transferred to the crossbar interface to be executed on real hardware or a simulator (see \Cref{fig:overview_sw}).

\vspace{-0.4cm}
\subsubsection{BNNs - From Integer to Digital Crossbar Arithmetic:}
For the translation of \acp{bnn} from integer to crossbar arithmetic, the following variables are used:
\begin{itemize}[topsep=1pt]
\item BNN's inputs/weights: $i_\text{NN}, w_\text{NN} \in \{-1, +1\}$
\item Digital crossbar inputs/weights: $v_\text{D}, g_\text{D} \in \{0, 1\}$
\end{itemize}

\begin{table}[!tp]
\centering
\footnotesize
\caption{Mapping of \ac{bnn} arithmetic to digital (D) crossbar arithmetic.}
\begin{tabular}{c||c|c|c|c}
Mapping & Approach & Equation: $o_\text{NN} = \sum_0^{N-1} i_\text{NN} w_\text{NN}$ & \#Cycles & \begin{tabular}[c]{@{}c@{}}\#Cells/\\ weight\end{tabular} \\ \hline\hline
BNN \Circled{I} & \begin{tabular}[c]{@{}c@{}}$ i_\text{NN} = 2 \cdot v_\text{D} - 1$ \\ $w_\text{NN}=g_\text{D}^+ - g_\text{D}^-$ \end{tabular} & $= 2 \left( \sum v_\text{D} g_\text{D}^+ - \sum v_\text{D} g_\text{D}^-  \right) - \sum w_\text{NN}$ & 1 & 2 \\ \hline
BNN \Circled{II} & \begin{tabular}[c]{@{}c@{}} $ i_\text{NN} = -2 \cdot v_\text{D} + 1$ \\ $ w_\text{NN}=g_\text{D}^+ - g_\text{D}^-$ \end{tabular} & $= 2 \left( \sum v_\text{D} g_\text{D}^- - \sum v_\text{D} g_\text{D}^+ \right) + \sum w_\text{NN}$ & 1 & 2 \\ \hline
\multirow{2}{*}{BNN \Circled{III}} & $i_\text{NN}=v_\text{D}^+ - v_\text{D}^-$  & \multirow{2}{*}{$= 2 \left( \sum v_\text{D}^+ g_\text{D} - \sum  v_\text{D}^- g_\text{D} \right) - \sum i_\text{NN}$} & 1 & 2 \\
                                & $w_\text{NN}=2\cdot g_\text{D} - 1$ &  & 2 & 1 \\ \hline
\multirow{2}{*}{BNN \Circled{IV}} & $i_\text{NN}=v_\text{D}^+ - v_\text{D}^-$ & \multirow{2}{*}{$= 2 \left( \sum v_\text{D}^- g_\text{D}  - \sum v_\text{D}^+ g_\text{D} \right) + \sum i_\text{NN}$} & 1 & 2 \\
                                &  $w_\text{NN}=-2\cdot g_\text{D} + 1$ &  & 2 & 1 \\ \hline
BNN \Circled{V} & XNOR & $= 2 \left( \sum v_\text{D}^+ g_\text{D}^+ + v_\text{D}^- g_\text{D}^- \right) - N $ & 1 & 2 \\ \hline
\multirow{2}{*}{BNN \Circled{VI}} & $i_\text{NN}=v_\text{D}^+ - v_\text{D}^-$ & \multirow{2}{*}{$= \sum v_\text{D}^+ g_\text{D}^+ + v_\text{D}^- g_\text{D}^- - v_\text{D}^+ g_\text{D}^- - v_\text{D}^- g_\text{D}^+ $} & 1 & 4 \\
                                    & $w_\text{NN}=g_\text{D}^+ - g_\text{D}^-$ &  & 2 & 2 \\

\end{tabular}
\label{tab:bnn_approaches}
\vspace{-0.4cm}
\end{table}

As mentioned before, the linear-scaling or differential mode can be used to omit negative inputs and weights.  
The usage of these modes can be chosen individually for both inputs and weights.  
These combinations lead to the possible mappings listed in \Cref{tab:bnn_approaches}.  
The column \textit{\#~Cycles} indicates the number of \acp{mvm} required to compute one \ac{mvm} in integer arithmetic.  
The column \textit{\#~Cells per weight} indicates the number of \acp{rram} cells needed per weight.  

For some mappings, e.g., \Circled{VI}, two realizations are possible:
Either more cycles or more cells per weight are needed.
The difference between \Circled{I+II} and \Circled{III+IV} is that in \Circled{I+II}, an offset of $\mp \sum g_\text{D}^+-g_\text{D}^-$ is added, which is known at compile time.
In \Circled{III+IV}, the offset $\mp \sum v_\text{D}^+-v_\text{D}^-$ depends on the inputs, i.e.,
the offset is not known at compile time.
This is why these variants require hardware support for adding the inputs.
The approach~\Circled{V} can be implemented using an \textit{XNOR} operation and is therefore also interesting for conventional hardware.

\vspace{-0.4cm}
\subsubsection{BNNs - From Digital to Analog Crossbar Arithmetic:}
To translate the mappings to analog crossbar arithmetic,
the following variables are used:
\begin{itemize}[topsep=1pt]
    \item Digital crossbar inputs: $v_\text{A} \in \{0, V_\text{r}\}, \quad V_\text{r} = V_\text{read}$
    \item Cell weights: $g_\text{A}^-, g_\text{A}^+ \in \{G_\text{min}, G_\text{max}\}, \quad G_\text{min} > 0 $
    \item Cell currents: $i_\text{A}^+, i_\text{A}^- \in \{I_\text{hrs}, I_\text{lrs}\}, \quad I_\text{hrs} > 0$
\end{itemize}

The mapping $v_\text{D} \rightarrow v_\text{A}$ is straightforward, as it can be mapped using the proportional relationship $v_\text{D}~=~v_\text{A}/V_\text{r}$, with the read voltage $V_\text{r}$.
More challenging is the mapping $g_\text{D} \rightarrow g_\text{A}$, as $G_\text{min} > 0$ leads to a non-zero offset:
\begin{itemize}[topsep=1pt]
    \item $g_\text{D} = \frac{g_\text{A} - G_\text{min}}{G_\text{mm}}$
    \item $g_\text{D}^{\{+,-\}} = \frac{g_\text{A}^{\{+,-\}} - G_\text{min}}{G_\text{mm}}, \quad G_\text{mm} \coloneqq G_\text{max} - G_\text{min}$
    \item $i_\text{D}^{\{+,-\}} = \frac{i_\text{A}^{\{+,-\}} - I_\text{hrs}}{I_\text{mm}}, \quad I_\text{mm} \coloneqq I_\text{lrs} - I_\text{hrs}$
\end{itemize}

\begin{table}[!tbp]
    \centering
    \footnotesize
    \caption{Mapping of \ac{bnn} arithmetic to analog crossbar arithmetic.}
    \vspace{0.1cm}
    \begin{tabular}{c||c|c|c}
    Mapping & Crossbar MVM(s) & Digital Correction & Analog Correction \\ \hline\hline
    BNN \Circled{I} & $\frac{2}{I_\text{mm}} \sum v_\text{A} \cdot (g_\text{A}^+-g_\text{A}^-)$ & $- \sum w_\text{NN}$ & / \\ \hline
    BNN \Circled{II} & $\frac{2}{I_\text{mm}} \sum v_\text{A} \cdot (g_\text{A}^--g_\text{A}^+)$ & $+ \sum w_\text{NN}$ & / \\ \hline
    BNN \Circled{III} & $\frac{2}{I_\text{mm}} \sum v_\text{A}^+ g_\text{A} - v_\text{A}^- g_\text{A}$ & $- \sum i_\text{NN}$ & $-2 \frac{I_\text{hrs}}{I_\text{mm}} \sum{i_\text{NN}}$ \\ \hline
    BNN \Circled{IV} & $\frac{2}{I_\text{mm}} \sum v_\text{A}^- g_\text{A} - v_\text{A}^+ g_\text{A}$ & $ \sum i_\text{NN}$ & $2 \frac{I_\text{hrs}}{I_\text{mm}} \sum{i_\text{NN}}$ \\ \hline
    BNN \Circled{V} & $\frac{2}{I_\text{mm}} \sum v_\text{A}^+ g_\text{A}^+ + v_\text{A}^- g_\text{A}^-$ & $- N$ & $-2 \frac{I_\text{hrs}}{I_\text{mm}} N$ \\ \hline
    \multirow{2}{*}{BNN \Circled{VI}} & $\frac{1}{I_\text{mm}} \sum v_\text{A}^+ g_\text{A}^+ + v_\text{A}^- g_\text{A}^-$  & \multirow{2}{*}{/} &  \multirow{2}{*}{/} \\
         & $\quad - v_\text{A}^+ g_\text{A}^- - v_\text{A}^- g_\text{A}^+$  & & \\
    \end{tabular}
    \label{tab:bnn_analog_mapping_results}
    \vspace{-0.5cm}
\end{table}

\vspace{0.1cm}
After applying these formulas to the equations in \Cref{tab:bnn_approaches}, the results listed in \Cref{tab:bnn_analog_mapping_results} can be observed.
The final equations for $o_\text{NN}$ can be obtained by adding the crossbar, digital correction, and analog correction terms:
\vspace{-0.2cm}
\begin{align}
o_\text{NN} = \mathrm{Crossbar \ MVM(s) + Digital \ Correction + Analog \ Correction}
\label{eq:bnn_to_analog_complete}
\end{align}
The first term (crossbar MVM(s)) is supposed to be executed on the crossbar.  
The digital correction term is a compile-time constant that results from the conversion of \ac{bnn} to digital crossbar arithmetic.  
The analog correction term is caused by the conversion of digital crossbar arithmetic to analog crossbar arithmetic.  
This term can only be omitted if $I_\text{hrs} \approx 0 ~ (G_\text{min} = 0)$ or $I_\text{mm} \gg I_\text{hrs}$.  
The column currents are converted to the digital domain by an \ac{adc}.

\vspace{-0.4cm}
\subsubsection{TNN Mappings:}
\label{sec:tnn_mappings}
\ac{tnn} arithmetic has three states for inputs and weights.
To map \acp{tnn} to crossbars with \SI{1}{\bit} weigths and inputs,
one can either split $i_\text{NN}$ and $w_\text{NN}$ again into their negative and positive parts, or represent inputs and weights as \SI{2}{\bit} binary values.
This requires two cells per \SI{2}{\bit} weight or two cycles per \SI{2}{\bit} input.
We use the notation $v_\text{D}=(v_\text{D}^1,v_\text{D}^0)$ for a \SI{2}{\bit} input and $g_\text{D}=(g_\text{D}^1,g_\text{D}^0)$ for a \SI{2}{\bit} weight.
\Cref{tab:tnn_approaches} shows five different mapping approaches to map \ac{tnn} arithmetic to digital crossbar arithmetic.
In comparison to the \ac{bnn} mapping, more cells per weight and/or more cycles are needed when using binary crossbars.
To obtain analog arithmetic, the same equations as for \acp{bnn} apply.

\begin{table}[!bp]
    \centering
    \footnotesize
    \vspace{-0.8cm}
    \caption{Mapping of \ac{tnn} arithmetic to digital crossbar arithmetic.}
    \begin{tabular}{c||c|l|c|c}
    Mapping & Approach & Equation: $o_\text{NN} = \sum_0^{N-1} i_\text{NN} w_\text{NN}$ & \#Cycles & \begin{tabular}[c]{@{}c@{}}\#Cells/\\ weight\end{tabular} \\ \hline\hline
    \multirow{2}{*}{TNN \Circled{I}} & $i_\text{NN}=v_\text{D}^+ - v_\text{D}^-$  & $= \sum g_\text{D}^+ v_\text{D}^+ + \sum g_\text{D}^- v_\text{D}^-$ & 2 & 2 \\
                                & $ w_\text{NN}=g_\text{D}^+ - g_\text{D}^-$ & $- \sum g_\text{D}^+ v_\text{D}^- - \sum g_\text{D}^- v_\text{D}^+ $ & 1 & 4 \\ \hline
    \multirow{2}{*}{TNN \Circled{II}} & $i_\text{NN}=(v_\text{D}^1,v_\text{D}^0)$  & $= \sum g_\text{D}^+ v_\text{D}^0 - \sum g_\text{D}^- v_\text{D}^0 $ & 2 & 2 \\
                                & $ w_\text{NN}=g_\text{D}^+ - g_\text{D}^-$ & $ - \left( \sum g_\text{D}^+ v_\text{D}^1 - \sum g_\text{D}^- v_\text{D}^1 \right) \ll 1$ & 1 & 4 \\ \hline
    \multirow{2}{*}{TNN \Circled{III}} & $i_\text{NN} + 1 = (v_\text{D}^1,v_\text{D}^0)$  & $= - \sum w_\text{NN} + \sum g_\text{D}^+ v_\text{D}^0 - \sum g_\text{D}^- v_\text{D}^0 $ & 2 & 2 \\
                                & $ w_\text{NN}=g_\text{D}^+ - g_\text{D}^-$ & $+ \left( \sum g_\text{D}^+ v_\text{D}^1 - \sum g_\text{D}^- v_\text{D}^1 \right) \ll 1$ & 1 & 4 \\ \hline
    \multirow{2}{*}{TNN \Circled{IV}} & $i_\text{NN} = v_\text{D}^+ - v_\text{D}^-$  & $= \sum g_\text{D}^0 v_\text{D}^+ - \sum g_\text{D}^0 v_\text{D}^- $ & 2 & 2 \\
                                & $ w_\text{NN}= (g_\text{D}^1, g_\text{D}^0)$ & $- \left( \sum g_\text{D}^1 v_\text{D}^+ - \sum g_\text{D}^1 v_\text{D}^- \right) \ll 1$ & 1 & 4 \\ \hline
    \multirow{2}{*}{TNN \Circled{V}} & $i_\text{NN} = v_\text{D}^+ - v_\text{D}^-$  & $= - \sum i_\text{NN} + \sum g_\text{D}^0 v_\text{D}^+ - \sum g_\text{D}^0 v_\text{D}^- $ & 2 & 2 \\
                                & $ w_\text{NN} + 1 = (g_\text{D}^1, g_\text{D}^0)$ & $+ \left( \sum g_\text{D}^1 v_\text{D}^+ - \sum g_\text{D}^1 v_\text{D}^- \right) \ll 1$ & 1 & 4 \\
    \end{tabular}
    \label{tab:tnn_approaches}
    \vspace{-0.0cm}
\end{table}

\vspace{-0.4cm}
\subsubsection{ADC Integration:}
\label{sec:adc}
The \ac{adc} is part of the simulator or hardware (see \Cref{fig:overview_sw}).
In this work, we use a simplified \ac{adc} model characterized by the input range $ADC_\text{range}\in[ADC_\text{in,min},ADC_\text{in,max}]$ and resolution $B$ in bits.
This approach allows us to effectively model clipping and quantization errors without requiring a fully analog-accurate \ac{adc} model.
\textit{Clipping errors} occur when the input signal exceeds the \ac{adc}'s input range.
Any input value outside the range is ``clipped'' to the maximum or minimum measurable value.
Quantization errors arise from the discretization process in analog-to-digital conversion.
Since the ADC can only represent the input with a finite number of discrete levels, there is an information loss between the true input signal and the digital counterpart.

When using differential weights, we assume that the corresponding columns are subtracted in the analog domain and then converted using the \ac{adc}.
This means, e.g., for mapping BNN \Circled{I} in \Cref{tab:bnn_analog_mapping_results}:
\vspace{-0.3cm}
\begin{align}
o_\text{NN} = \frac{2}{I_\text{mm}} ADC \left( \sum (i_\text{A}^+-i_\text{A}^-) \right) - \sum w_\text{NN} \label{eq:analog_1_adc_B}
\end{align}
\vspace{-0.5cm}

In addition to range and resolution, we introduce the clipping factor $\alpha$.
This factor specifies the proportion of the maximum input range that is utilized.
For differential mappings, the maximum possible input is $i_\text{max} = N \cdot (I_\text{lrs} - I_\text{hrs})$.
This results in the following equation:
\vspace{-0.2cm}
\begin{align}
    ADC\left( x \right) &=  sgn(x) \Delta  \left(  \left\lfloor \frac{clip \left( |x|, \ - \alpha \cdot i_{\text{max},B}, \ \alpha \cdot i_{\text{max},B} \right)}{\Delta} \right\rfloor + \frac{1}{2} \right) \label{eq:quant_ADC_B}
\end{align}
\vspace{-0.4cm}

The output from \Cref{eq:quant_ADC_B} is again a current that includes clipping and quantization errors.
The function $Q(x)= sgn(x)\Delta \left( \lfloor |x|/ \Delta \rfloor + 1/2 \right) $ represents the general quantization function of a mid-rise quantizer for a signed input signal $x$.
The step width for $B$ bit resolution is $\Delta=\alpha\cdot 2i_\text{max}/2^B$.

\vspace{-0.4cm}
\section{Results}
\label{sec:results}
\vspace{-0.3cm}
CIM-Explorer can be used to analyze the interplay between properties of \ac{rram} crossbars and the mapping strategy on the inference accuracy of \acp{bnn} and \acp{tnn}.  
To demonstrate its capabilities, we exemplarily explore the effects of \ac{adc} parameters and cell variability in combination with different mapping techniques.
The crossbar size used is $256{\times}256$, as this offers a good compromise between cell utilization and energy efficiency~\cite{cubero2025evaluating}.

\begin{figure*}[t!]
    \centering
    \begin{subfigure}[b]{0.28\linewidth}
        \centering
        \includegraphics[width=\linewidth, keepaspectratio]{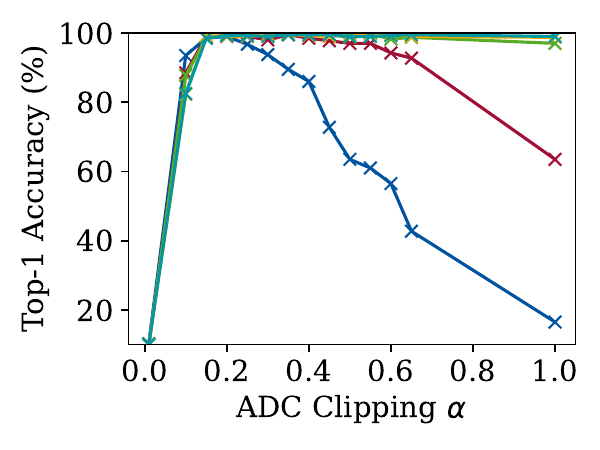}
        \vspace{-0.6cm}
        \caption{BNN \Circled{I}}
        \label{fig:adc_res_bnn_I}
    \end{subfigure}
    \begin{subfigure}[b]{0.28\linewidth}
        \centering
        \includegraphics[width=\linewidth, keepaspectratio]{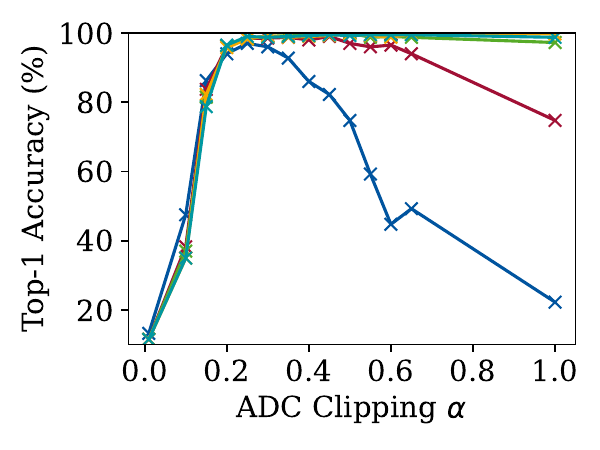}
        \vspace{-0.6cm}
        \caption{BNN \Circled{II}}
        \label{fig:adc_res_bnn_II}
    \end{subfigure}
    \begin{subfigure}[b]{0.28\linewidth}
        \centering
        \includegraphics[width=\linewidth, keepaspectratio]{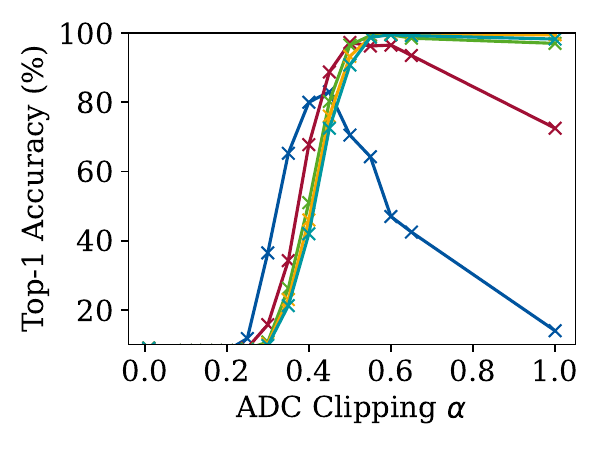}
        \vspace{-0.6cm}
        \caption{BNN \Circled{III}}
        \label{fig:adc_res_bnn_III}
    \end{subfigure}

    \vspace{0.0cm}
    \begin{subfigure}[b]{0.28\linewidth}
        \centering
        \includegraphics[width=\linewidth, keepaspectratio]{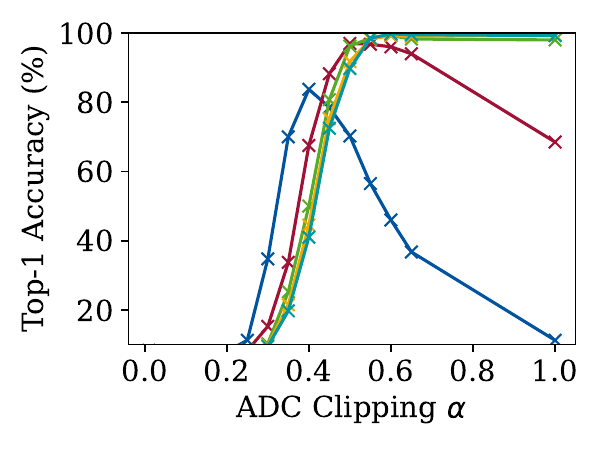}
        \vspace{-0.6cm}
        \caption{BNN \Circled{IV}}
        \label{fig:adc_res_bnn_IV}
    \end{subfigure}
    \begin{subfigure}[b]{0.28\linewidth}
        \centering
        \includegraphics[width=\linewidth, keepaspectratio]{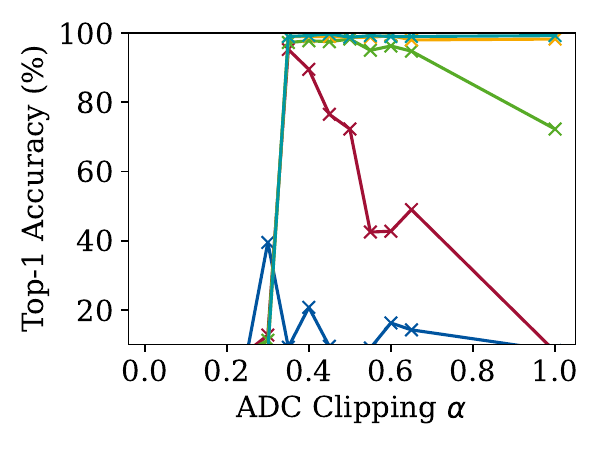}
        \vspace{-0.6cm}
        \caption{BNN \Circled{V}}
        \label{fig:adc_res_bnn_V}
    \end{subfigure}
    \begin{subfigure}[b]{0.28\linewidth}
        \centering
        \includegraphics[width=\linewidth, keepaspectratio]{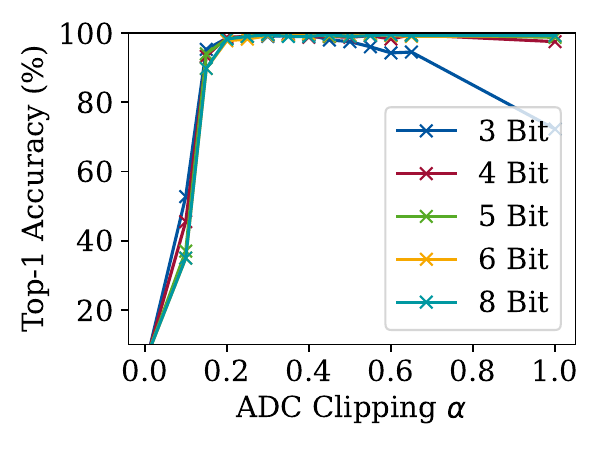}
        \vspace{-0.6cm}
        \caption{BNN \Circled{VI}}
        \label{fig:adc_res_bnn_VI}
    \end{subfigure}
    \vspace{-0.2cm}
    \caption{Top-\SI{1}{\percent} classification accuracy for CIFAR-10 (train set) on VGG-7 for different \ac{adc} resolutions and \ac{bnn} mappings depending on parameter $\alpha_\text{ADC}$.}
    \label{fig:adc_res_all}
    \vspace{-0.0cm}
\end{figure*}

\vspace{-0.4cm}
\subsubsection{ADC Impact}
The \ac{adc} consumes a significant portion of the power in \ac{rram} crossbars~\cite{shafiee2016isaac}.
Therefore, a low resolution is advantageous, but clipping and quantization errors reduce accuracy.
To analyze these trade-offs, we assume an infinite \ac{adc} resolution.
The parameters are set to $I_\text{hrs} = \SI{5}{\micro\ampere}$ and $I_\text{lrs} = \SI{10}{\micro\ampere}$.

\Cref{fig:adc_res_all} shows the Top-\SI{1}{\percent} classification accuracy for CIFAR-10 trained on VGG-7 across different \ac{bnn} mappings, \ac{adc} resolutions, and clipping factors $\alpha$.  
The \ac{bnn} \Circled{VI} mapping achieves the highest accuracy but also requires the most cells per weight (see \Cref{sec:intvscrossbararith}).  
Even at an \ac{adc} resolution of just \SI{3}{\bit}, the original accuracy is maintained over a wide range of $\alpha$.  
Mappings \ac{bnn} \Circled{I+II} and \Circled{III+IV} show similar results.  
This is expected since these mappings differ only in weight sign and correction factors.  
\ac{bnn} \Circled{I} and \Circled{II} outperform \ac{bnn} \Circled{III} and \Circled{IV} and should therefore be preferred.  
Additionally, they eliminate the need for an analog correction term.  
Although the XNOR mapping (\ac{bnn} \Circled{V}) is often used in the literature~\cite{sun2018xnor}, it performs the worst and offers no advantages over \ac{bnn} \Circled{I} and \Circled{II} (see \Cref{tab:bnn_approaches}).  
\textit{CIM-Explorer makes it possible to compare these mappings and select the best one for a given crossbar.  
Furthermore, the required \ac{adc} parameters can be determined.}

\begin{figure*}[t!]
    \centering
    \vspace{-0.1cm}
    \captionsetup[subfigure]{justification=centering}
    \begin{subfigure}[b]{0.25\linewidth}
        \centering
        \includegraphics[width=\linewidth, keepaspectratio]{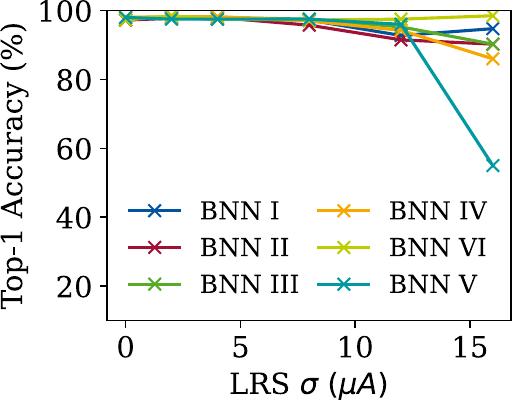}
        \vspace{-0.5cm}
        \caption{$\mu_\text{hrs}=\SI{5}{\micro\ampere}$}
        \label{fig:hrs5_lrs30_LRS_SIGMA}
    \end{subfigure}
    \begin{subfigure}[b]{0.23\linewidth}
        \centering
        \includegraphics[width=\linewidth, keepaspectratio]{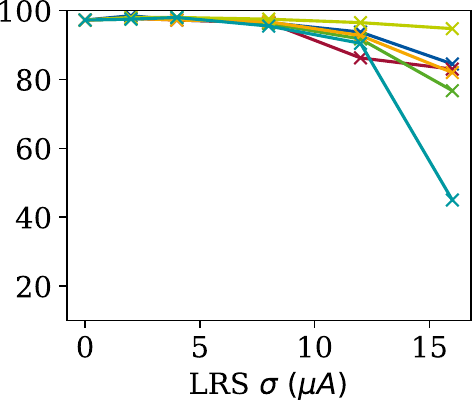}
        \vspace{-0.5cm}
        \caption{$\mu_\text{hrs}=\SI{10}{\micro\ampere}$}
        \label{fig:hrs10_lrs30_LRS_SIGMA}
    \end{subfigure}
    \begin{subfigure}[b]{0.23\linewidth}
        \centering
        \includegraphics[width=\linewidth, keepaspectratio]{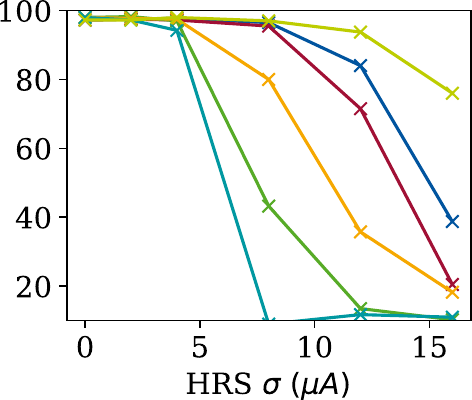}
        \vspace{-0.5cm}
        \caption{$\mu_\text{hrs}=\SI{5}{\micro\ampere}$}
        \label{fig:hrs5_lrs30_HRS_SIGMA}
    \end{subfigure}
    \begin{subfigure}[b]{0.23\linewidth}
        \centering
        \includegraphics[width=\linewidth, keepaspectratio]{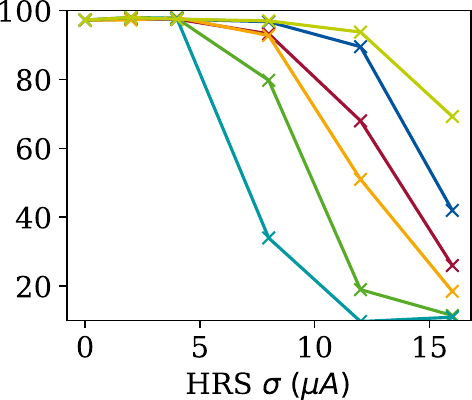}
        \vspace{-0.5cm}
        \caption{$\mu_\text{hrs}=\SI{10}{\micro\ampere}$}
        \label{fig:hrs10_lrs30_HRS_SIGMA}
    \end{subfigure}
    \vspace{-0.3cm}
    \caption{BNN accuracy for CIFAR-10 (train set) on VGG-7 for different LRS/HRS placements and different cell variabilites.
    $\mu_\text{lrs}$ is $\SI{30}{\micro\ampere}$ and $M_\text{int}=N_\text{int}=256$.}
    \label{fig:var_res_all}
    \vspace{-0.5cm}
\end{figure*}

\vspace{-0.5cm}
\subsubsection{\ac{lrs} and \ac{hrs} Variability}
Cell variability also reduces the inference accuracy, which will be further examined in the next experiment.  
We model state variability using a normal distribution with means $\mu_\text{L} = I_\text{lrs}$ and $\mu_\text{H} = I_\text{hrs}$, and standard deviations $\sigma_\text{lrs}$ and $\sigma_\text{hrs}$.  
To isolate the effects of variability, we assume an infinitely high \ac{adc} resolution.

\Cref{fig:hrs5_lrs30_LRS_SIGMA,fig:hrs10_lrs30_LRS_SIGMA} show the accuracy for different mappings depending on $\sigma_\text{lrs}$.
\Cref{fig:hrs5_lrs30_HRS_SIGMA,fig:hrs10_lrs30_HRS_SIGMA} show the accuracy depending on $\sigma_\text{hrs}$.  
The mapping \ac{bnn} \Circled{VI} performs best, while \ac{bnn} \Circled{V} performs worst in terms of tolerance to cell variability.  
At first glance, \ac{hrs} variability appears to have a greater impact on accuracy than \ac{lrs} variability.  
However, this effect also stems from the modeling approach:  
Since negative currents are not possible, the Gaussian distribution around $I_\text{hrs}$ is asymmetric, which contributes to the strong accuracy drop.  
It becomes clear that the variability distributions, the state location, and the mapping impact accuracy.  
\textit{Their interplay is highly complex, requiring tools like ours to simulate various scenarios and to choose the best mapping.}

\vspace{-0.4cm}
\subsubsection{Large-Size BNNs}
The following experiments show how the previous findings scale to larger BNNs and datasets.
Therefore, we train BinaryNet, BinaryDenseNet28, and BinaryDenseNet37 on CIFAR-100.
Since the \ac{bnn} \Circled{VI} mapping achieves the highest accuracy under non-idealities, this mapping is used.
The first row of \Cref{tab:big_bnns_results} presents the test accuracy, while the remaining columns report the best results from experiments where the absolute accuracy drop remains below \SI{1}{\percent} compared to the baseline.
For BinaryDenseNet37, an \ac{adc} resolution of just \SI{3}{\bit} suffices, like in the much smaller VGG-7 model.
The variability results show that larger BNNs trained on the same dataset tend to be less sensitive to non-linearities.
\textit{
CIM-Explorer shows that smaller \acp{nn} with similar baseline accuracy should not always be preferred over larger models, as larger models might achieve higher accuracy under non-idealities on \ac{rram} crossbars.
}

\begin{table}[!tbp]
\centering
\caption{Maximum tolerable non-idealities for larger \acp{bnn} trained on CIFAR-100. Mapping BNN \Circled{VI} is used with $\mu_\text{hrs}=~$\SI{5}{\micro\ampere} and $\mu_\text{lrs}=~$\SI{30}{\micro\ampere}.}
\vspace{0.2cm}
\begin{tabular}{c||c|c|c}
                   & BinaryNet & BinaryDenseNet28 & BinaryDenseNet37 \\ \hline \hline
Top-\SI{1}{\percent} (Test Set) Accuracy & \SI{45.3}{\percent} & \SI{88.0}{\percent} & \SI{88.1}{\percent} \\ \hline
Minimum ADC resolution & \SI{4}{\bit} & \SI{4}{\bit}  &  \SI{3}{\bit} \\ \hline
Maximum LRS sigma (\SI{}{\micro\ampere}) & {$\sigma_\text{lrs}=~$\SI{4}{\micro\ampere}} &  {$\sigma_\text{lrs}=$~\SI{5}{\micro\ampere}} &  {$\sigma_\text{lrs}=$~\SI{8}{\micro\ampere}}  \\ \hline
Maximum HRS sigma (\SI{}{\micro\ampere}) &  $\sigma_\text{hrs}=~$\SI{5}{\micro\ampere}   &  $\sigma_\text{hrs}=$~\SI{5}{\micro\ampere}  &   $\sigma_\text{hrs}=$~\SI{5}{\micro\ampere}  \\
\end{tabular}
\label{tab:big_bnns_results}
\vspace{-0.5cm}
\end{table}

\begin{figure*}[b!]
    \centering
    \vspace{-0.6cm}
    \begin{subfigure}[b]{0.28\linewidth}
        \centering
        \includegraphics[width=\linewidth, keepaspectratio]{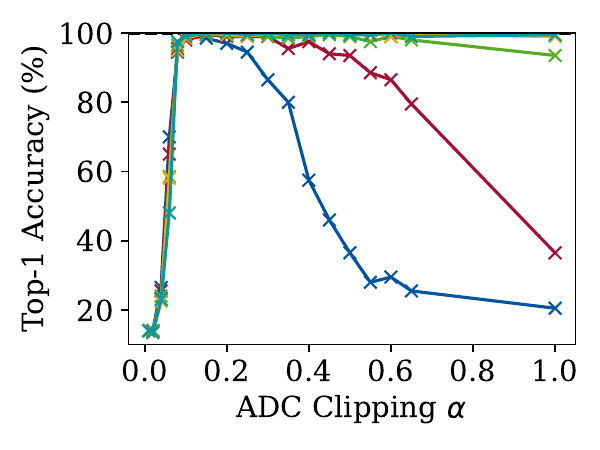}
        \vspace{-0.6cm}
        \caption{TNN \Circled{I}}
        \label{fig:adc_res_tnn_I}
    \end{subfigure}
    \begin{subfigure}[b]{0.28\linewidth}
        \centering
        \includegraphics[width=\linewidth, keepaspectratio]{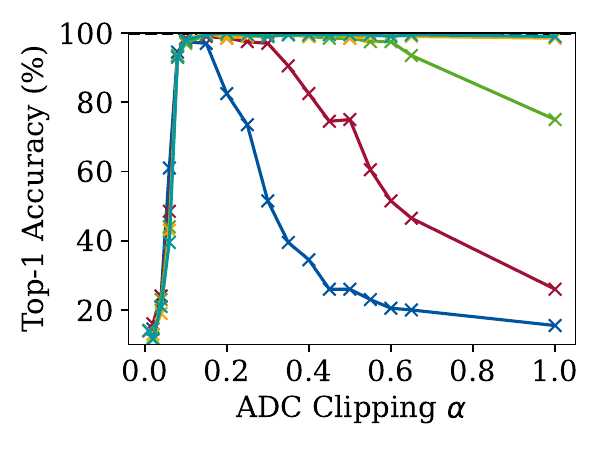}
        \vspace{-0.6cm}
        \caption{TNN \Circled{II}}
        \label{fig:adc_res_tnn_II}
    \end{subfigure}
    \begin{subfigure}[b]{0.28\linewidth}
        \centering
        \includegraphics[width=\linewidth, keepaspectratio]{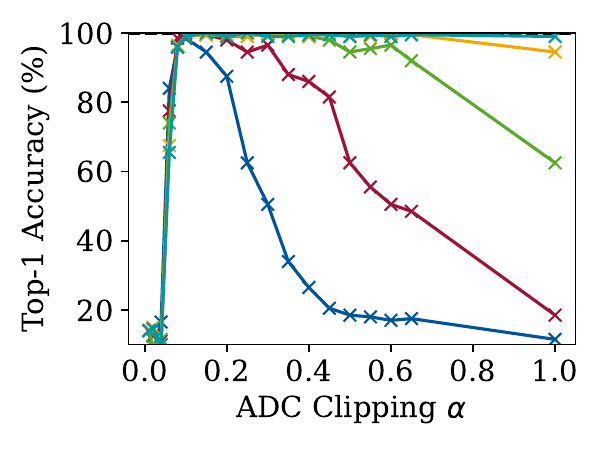}
        \vspace{-0.6cm}
        \caption{TNN \Circled{III}}
        \label{fig:adc_res_tnn_III}
    \end{subfigure}

    \vspace{0.0cm}
    \begin{subfigure}[b]{0.28\linewidth}
        \centering
        \includegraphics[width=\linewidth, keepaspectratio]{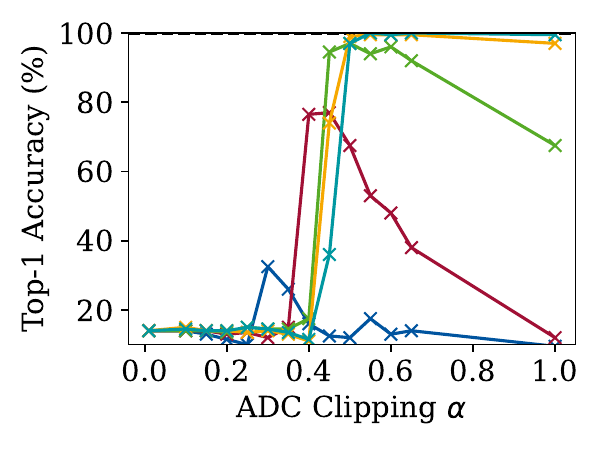}
        \vspace{-0.4cm}
        \caption{TNN \Circled{IV}}
        \label{fig:adc_res_tnn_IV}
    \end{subfigure}
    \begin{subfigure}[b]{0.28\linewidth}
        \centering
        \includegraphics[width=\linewidth, keepaspectratio]{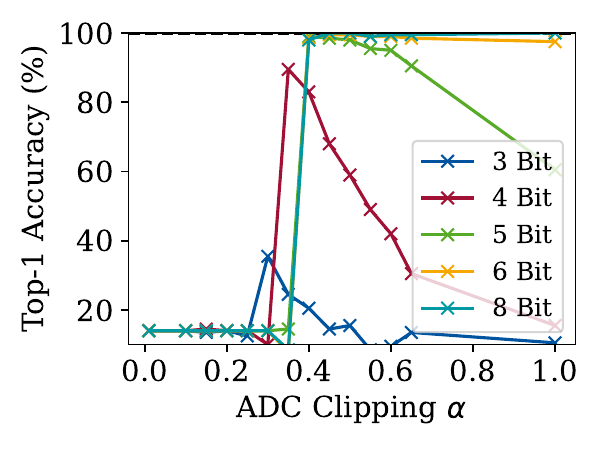}
        \vspace{-0.4cm}
        \caption{TNN \Circled{V}}
        \label{fig:adc_res_tnn_V}
    \end{subfigure}
    \vspace{-0.2cm}
    \caption{Top-\SI{1}{\percent} classification accuracy for CIFAR-10 (train set) on VGG-7 for different \ac{adc} resolutions and \ac{tnn} mappings depending on parameter $\alpha_\text{ADC}$.}
    \label{fig:adc_tnn_res_all}
\end{figure*}

\vspace{-0.4cm}
\subsubsection{Comparison to TNNs}
Finally, we show the TNN results and compare them to the BNN mappings.
$I_\text{hrs}$ is \SI{5}{\micro\ampere}, and $I_\text{lrs}$ is \SI{10}{\micro\ampere}.
\Cref{fig:adc_tnn_res_all} presents accuracy results for TNN mappings \Circled{I} to \Circled{V} across different \ac{adc} resolutions.
Overall, the differential weight mappings \Circled{I-III} outperform the linear-scaling mappings \Circled{IV+V}.
TNN \Circled{I} should be preferred over \Circled{II+III} when supported by the hardware (see \Cref{sec:tnn_mappings}).
Among the linear-scaling mappings, which require fewer cells, TNN \Circled{V} achieves higher accuracy than \Circled{IV}, however, its correction term is more complex (see \Cref{tab:tnn_approaches}).
Since \acp{tnn} contain more zeros than \acp{bnn} after the digital mapping,
the clipping factors are in general lower than with \acp{bnn}.
BNN \Circled{VI} has a wider acceptable range of $\alpha$ at \SI{3}{\bit} resolution.
Hence, it slightly outperforms the TNN mappings.
However, TNNs can generally be trained to higher accuracy than BNNs~\cite{zhu2022tab}.

\begin{figure*}[t!]
    \centering
    \captionsetup[subfigure]{justification=centering}
    \begin{subfigure}[b]{0.25\linewidth}
        \centering
        \includegraphics[width=\linewidth, keepaspectratio]{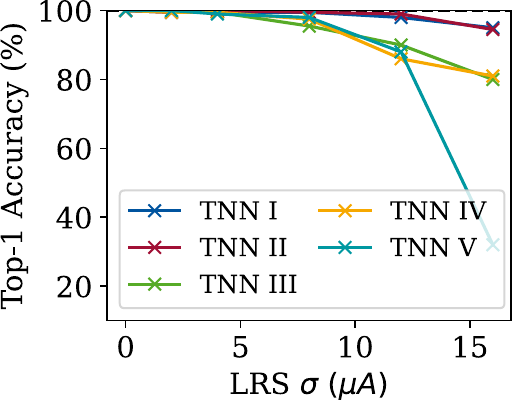}
        \vspace{-0.4cm}
        \caption{$\mu_\text{hrs}=\SI{5}{\micro\ampere}$}
        \label{fig:tnn_hrs5_lrs30_LRS_SIGMA}
    \end{subfigure}
    \begin{subfigure}[b]{0.23\linewidth}
        \centering
        \includegraphics[width=\linewidth, keepaspectratio]{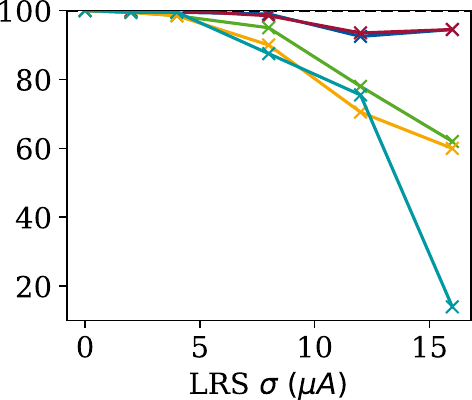}
        \vspace{-0.4cm}
        \caption{$\mu_\text{hrs}=\SI{10}{\micro\ampere}$}
        \label{fig:tnn_hrs10_lrs30_LRS_SIGMA}
    \end{subfigure}
    \begin{subfigure}[b]{0.23\linewidth}
        \centering
        \includegraphics[width=\linewidth, keepaspectratio]{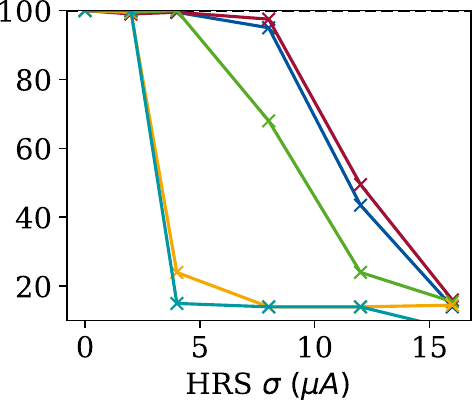}
        \vspace{-0.4cm}
        \caption{$\mu_\text{hrs}=\SI{5}{\micro\ampere}$}
        \label{fig:tnn_hrs5_lrs30_HRS_SIGMA}
    \end{subfigure}
    \begin{subfigure}[b]{0.23\linewidth}
        \centering
        \includegraphics[width=\linewidth, keepaspectratio]{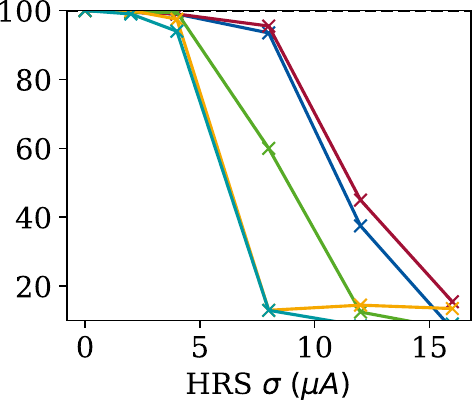}
        \vspace{-0.4cm}
        \caption{$\mu_\text{hrs}=\SI{10}{\micro\ampere}$}
        \label{fig:tnn_hrs10_lrs30_HRS_SIGMA}
    \end{subfigure}
    \vspace{-0.2cm}
    \caption{TNN accuracy for CIFAR-10 (train set) on VGG-7 for different LRS/HRS placements and different cell variabilites.
    $I_\text{lrs}$ is $\SI{30}{\micro\ampere}$ and $M_\text{int}=N_\text{int}=256$.}
    \label{fig:var_tnn_res_all}
    \vspace{-0.5cm}
\end{figure*}

\Cref{fig:var_tnn_res_all} shows the Top-\SI{1}{\percent} accuracy under cell variability.
For LRS variability, mappings TNN \Circled{I+II} exhibit similar robust behavior as BNN \Circled{VI}, but TNNs are less robust against HRS variability.
This is likely because the value $0$, mapped to the HRS, occurs more frequently in TNNs than $\pm 1$.
In contrast, in the differential BNN mappings, LRS and HRS occur with equal frequency.  
\textit{These experiments demonstrate that our tool not only supports BNNs but also TNNs, without the need for additional modifications.}
\vspace{-0.3cm}

\section{Conclusion}
\label{sec:conclusion}
\vspace{-0.3cm}
In this paper, we presented CIM-Explorer, a modular toolkit for the exploration of \ac{bnn} and \ac{tnn} inference on RRAM crossbars.  
Our work integrates a compiler, various mapping techniques and simulators.
In the results, we demonstrated how CIM-Explorer can be used for a \ac{dse} at an early design stage.

CIM-Explorer not only provides a comprehensive framework to analyze the trade-offs associated with different mapping strategies and their impact on inference accuracy under varying non-idealities,  
but it also has a modular structure, allowing the individual components to be used separately.  
This means that our compiler can easily be used with real hardware, mappings can be exchanged, and other simulators can be integrated.
The research community can use our open-source toolkit to help advance the development of CIM technologies.

\vspace{-0.3cm}

\begin{footnotesize}
\subsubsection*{Acknowledgments:}
This work was funded by the German BMWK and the European Union ESF Plus fund under the grant number 03EFWNW338.
\end{footnotesize}

\vspace{-0.4cm}
\bibliographystyle{splncs04}
\bibliography{bibtexentry}

\end{document}